\documentclass[12pt]{article}
\usepackage{a4}
\usepackage{epsf}
\usepackage{amssymb}
\usepackage{cite}
\newcommand{\be}{\begin{equation}}
\newcommand{\ee}{\end{equation}}
\newcommand{\bea}{\begin{eqnarray}}
\newcommand{\eea}{\end{eqnarray}}

\begin{document}
\def\C{{\mathbb{C}}}
\def\R{{\mathbb{R}}}
\def\s{{\mathbb{S}}}
\def\T{{\mathbb{T}}}
\def\Z{{\mathbb{Z}}}
\def\W{{\mathbb{W}}}
\def\Bbb{\mathbb}
\def\BZ{\Bbb Z} \def\BR{\Bbb R}
\def\BW{\Bbb W}
\def\BM{\Bbb M}
\def\e{\mbox{e}}
\def\BC{\Bbb C} \def\BP{\Bbb P}
\def\CP{\BC\BP}
\begin{titlepage}
\title{On Cardy states in the (2,2,2,2) Gepner model}
\author{}
\date{
Gor Sarkissian
\thanks{\noindent E--mail:~ gor.sarkissian@roma2.infn.it}
 \vskip0.4cm
{Dipartimento di Fisica \\
Universit\`a di Roma ``Tor Vergata'',\\
I.N.F.N Sezione di Roma ``Tor Vergata'' \\
Via della Ricerca Scientifica 1, 00133 Roma, Italy}} \maketitle
\abstract{We study Cardy states in the (2,2,2,2) Gepner model from
both the algebraic and geometric vantage points. We present the
full list of primaries of this model together with their
characters. The effects of fixed point resolution are analyzed.
Annulus partition functions between various Cardy states are
calculated. Using the equivalent description in terms of the
$T^4/Z_4$ orbifold, the corresponding geometric realization is
partially found.}
\end{titlepage}

\section{Introduction and summary}\label{intro}

The latest developments in String Theory demonstrated the
importance of understanding properties of D-branes in curved
backgrounds. Despite widespread effort our knowledge of D-branes
properties is still limited to the simplest backgrounds, like tori
or toroidal orbifolds, group manifolds etc. It turned out that in
most applications of D-branes to string theory detailed
understanding of the D-branes on more complicated backgrounds,
first of all Calabi-Yau manifolds, is necessary. Unfortunately for
general Calabi-Yau manifolds not much is known on special
Lagrangian submanifolds or holomorphic cycles wrapped by A- and
B-type branes respectively.

However it is known that at certain points in their moduli spaces,
Calabi-Yau compactifications can be described by rational
conformal field theories, known as Gepner models
\cite{Gepner:1989gr}. By now a sophisticated technique for
constructing boundary states  on rational conformal field theories
exists
\cite{Cardy:1989ir,Cardy:1991tv,Sagnotti:1987tw,Pradisi:1988xd,Bianchi:1990yu,Bianchi:1990tb,Pradisi:1995qy,Pradisi:1995pp,Pradisi:1996yd,Bianchi:1991eu}.
Starting from the pioneering papers
\cite{Ooguri:1996ck,Recknagel:1997sb}, a number of papers
\cite{Gutperle:1998hb,Brunner:1999jq,Govindarajan:1999js,Naka:2000he,Scheidegger:1999ed,Kaste:1999id,
Diaconescu:1999vp,Fuchs:2000fd,Mizoguchi:2001xi,Brunner:2000nk,Fuchs:2000gv,Recknagel:2002qq,Brunner:2005fv,Enger:2005jk,Brunner:2006tc}
were devoted to boundary states in Gepner models. Much of the work
was devoted to figure out geometrical properties  of branes given
by  specific boundary states. These papers also brought to light
that in the presence of minimal models with even levels some of
the boundary states defined in \cite{Recknagel:1997sb} should be
modified or, better, resolved in order to dispose of fixed point
ambiguities \cite{Brunner:2000nk}, \cite{Fuchs:2000gv}. General
formulae for boundary states in conformal field theories with
simple current modular invariants have been given in
\cite{Fuchs:2000cm,Fuchs:2004dz}.

As usual before tackling complicated cases some simple models were
analyzed. In \cite{Gutperle:1998hb} the $(1,1,1)\sim(1,4)$ and
$(2,2)$ models, which are equivalent to $T^2$ compactifications at
the $SU(3)_1$ and $SO(4)_1$ enhancement points respectively, were
discussed. In \cite{Brunner:2006tc} B-branes in the $(2,2,2,2)$
model, which admits orbifold description as $T^4/Z_4$
\cite{Eguchi:1988vr},\cite{Nahm:1999ps}, were discussed.

In this paper we study in depth the Cardy states corresponding to
$D0$ branes in the $(2,2,2,2)$ model .

The paper is organized as follows.

In section \ref{rev} we review necessary background material on the simple current extensions.

In section \ref{genprop} we review Gepner models via simple current extension formalism.

In section \ref{prim} we write down all the necessary information on the $(2,2,2,2)$ model: orbit representatives,
characters, conformal weights.
 Using the resolved characters we compute the torus partition function and show
that it coincides with the one computed in the appendix C as an
orbifold partition function at the $SU(2)^4$ point. Using the
general formulae of section \ref{rev} we also derive the annulus
partition functions between different Cardy states, paying special
attention to the peculiarities caused by the presence of the fixed
points.

In section \ref{brorb} we study $D0$ branes on the orbifold
$T^4/Z_4$ . We compute all the annulus partition functions between
$D0$ branes located at points in $T^4/Z_4$ orbifold that are fully
or partially fixed under the orbifold group action. Using
previously derived formulae for the annulus partition functions
between Cardy states of the (2,2,2,2) model we establish a partial
dictionary between Cardy states  and $D0$ branes. This is the main
result of this paper.

The necessary formulae on theta functions are reviewed in appendices A and B.

\section{Simple current extension: brief review}\label{rev}

Let us briefly remind the meaning of the simple current extension
by simple currents of integral conformal weight
\cite{Schellekens:1990xy,Blumenhagen:1995tt,Fuchs:1996dd,Fuchs:1999zi,Fuchs:1999xn},
\cite{Fuchs:2000gv}. A primary $J$ is called a simple current if,
fused with any other primary $\lambda$, it yields just a single
field $J\lambda$. Simple current extension means the combination
of two operations:
\begin{itemize}
\item Projection. We keep only fields which obey $Q_J(\lambda)=0$
where \be\label{neut}
 Q_J(\lambda)=\Delta_{\lambda}+\Delta_{J}-\Delta_{J\lambda}\;\; ({\rm mod}\;\;\; Z)
 \ee
 \item
 Extension. We extend the chiral algebra by including the simple current $J$. This means that we organize the fields surviving the projection
 into orbits derived as a result of fusion with the simple current $J$.
\end{itemize}

Before writing the torus partition function we should discuss the
important issue of fixed point resolution. If all the primaries
form orbits of the same length, equal to the order $|G|$ of the
full group $G$ generated by the simple currents, or in other words
have the same number of images under the repeated fusion with the
simple current, the characters could be labelled by the primaries
chosen, one from each orbit, called orbit representatives, and
have the form: \be\label{charex1}
 \tilde{\chi}_{\hat{\lambda}}=\sum _{J\in G}\chi_{J\lambda}\ee

 The unitary matrix representing modular transformations on the extended theory is:
 \be\label{ssimp}
 \tilde{S}_{\hat{a},\hat{b}}=|G|S_{ab}
 \ee
 where with hatted variables we denoted the orbit representatives.
 However it may happen that some of the primaries have a
non-trivial stabilizer ${\cal S}_{\lambda}$, i.e. be fixed under
the action of currents of a subgroup ${\cal S}_{\lambda}\in G$. In
this case the freely acting group is the factor $G_a=G/{\cal S}_a$
and the orbit length is given by \be |G_a|=\frac{|G|}{|{\cal
S}_a|} \ee and therefore varies from orbit to orbit. The  simple
formula (\ref{ssimp}) for the modular transformation matrix does
not work anymore. It turns out that in order to construct a
unitary matrix representation of the modular transformation in
this case one needs to resolve the primaries with non-trivial
stabilizer, i.e. one should consider together with the orbit
$\hat{a}$ additional $|{\cal S}_a|$ orbits\footnote{Actually each
primary should be resolved by the order of the subgroup ${\cal
U}_a$ of the stabilizer, called untwisted stabilizer
\cite{Fuchs:1996dd}, on which a certain alternating $U(1)$-valued
bihomomorphism, or discrete torsion, on the stabilizer ${\cal
S}_a$ vanishes. It is well-known that discrete torsions are
classified by the second $U(1)$-valued cohomology group $H^2({\cal
S}_a, U(1))$ \cite{Vafa:1986wx}, and since in Gepner models with
diagonal (or charge conjugation) torus partition function -- the
situation of our interest below -- the stabilizers are all
isomorphic to the $Z_2$ group, for which $H^2(Z_2, U(1))=0$ , one
finds that the untwisted stabilizer coincides with stabilizer. }.
Labelling the additional orbits by $i$ we find the characters:
\be\label{charex}
 \tilde{\chi}_{\hat{\lambda},i}=m_{\hat{\lambda},i}\sum _{J\in G/{\cal S}_{\lambda}}\chi_{J\lambda}
=\frac{m_{\hat{\lambda},i}}{|{\cal S_{\lambda}}|}\sum _{J\in
G}\chi_{J\lambda}\ee where $m_{i,a}$ are usually equal to $1$, but
we keep them explicitly so as to keep track of the different
resolved orbits.

 The diagonal modular invariant torus partition function of the extended theory reads\be
 Z_{\rm ext}=\sum_{\hat{\lambda},i}|\tilde{\chi}_{\lambda,i}|^2=\sum_{\rm{orbits}\, Q(\lambda)=0}|{\cal S}_{\lambda}|\cdot |\sum_{J\in
G/{\cal S}_{\lambda}}\chi_{J\lambda}|^2\ee
where we used that
\be\label{ms} |{\cal S}_a|=\sum_i
(m_{a,i})^2 \ee
The unitary matrix representation of the modular transformation $S$ on the characters (\ref{charex}), was constructed in
\cite{Schellekens:1990xy}, \cite{Fuchs:1996dd}.
The following ansatz was suggested
\be\label{schyaa}
\tilde{S}_{(a,i),(b,j)}=m_{a,i}m_{b,j}{|G_a||G_b|\over
|G|}S_{a,b}+\Gamma_{(a,i),(b,j)}\ee
 where $\Gamma_{(a,i),(b,j)}$
satisfies the equation \be\label{nco2}
\sum_j\Gamma_{(a,i),(b,j)}m_{b,j}=0 \ee and it is therefore
different from zero only between fixed points. It was found in
\cite{Schellekens:1990xy} that unitarity requires
$\Gamma_{(a,i),(b,j)}$ to satisfy the condition: \be\label{uncon}
\sum_{\rm{orbits}\,
Q(b)=0,j}\Gamma_{(a,i),(b,j)}\Gamma_{(c,k),(b,j)}^*=\delta_{ac}(\delta_{ik}-{m_{a,i}m_{a,k}\over
|S_a|}) \quad .\ee

The derivation of equation (\ref{uncon}) is reviewed in appendix
D.

Using the matrix (\ref{schyaa}) one can compute the fusion rule
coefficients using Verlinde formula and the annulus partition
functions for the Cardy states. After some algebra, reviewed in
appendix D, we arrive at the expression:

\bea\label{anamppp} A_{(a,i),(d,e)}&=&\sum_{\rm{orbits}\,
Q(c)=0}\sum_{J\in G}{m_{a,i}m_{d,e}{\cal N}_{Ja,c}^d\over |{\cal
S}_a||{\cal S}_d|}\sum _{K\in G_c}\chi_{Kc}\nonumber\\
&+&\sum_{\rm{orbits}\, Q(c)=0}\sum_{(\rm{orbits}\,
Q(b)=0,j)}{\Gamma_{(a,i),(b,j)}S_{c,b}\Gamma_{(b,j),(d,e)}^*\over
S_{0,b}}\sum _{K\in G_c}\chi_{Kc} \eea Given that the resolving
matrix $\Gamma_{(a,i),(b,j)}$ are different from zero only between
fixed points we observe that formula (\ref{anamppp}) simplifies if
one of the states is not fixed. When $a$ is not fixed and $d$
fixed (\ref{anamppp}) simplifies to \be\label{anaff}
A_{(a),(d,e)}=\sum_{\rm{orbits}\, Q(c)=0}\sum_{J\in
G}{m_{d,e}{\cal N}_{Ja,c}^d\over |{\cal S}_d|}\sum _{K\in
G_c}\chi_{Kc} \ee When neither $a$ nor $d$ are fixed
(\ref{anamppp}) further simplifies to \bea\label{anag}
A_{ad}=\sum_{\rm{orbits}\, Q(c)=0}\sum_{J\in G}{\cal N}_{Ja,c}^d
\sum _{K\in G_c}\chi_{Kc} \eea For later application to Gepner
models let us discuss the matrix $\Gamma_{(a,i),(b,j)}$ and the
second term in (\ref{anamppp}) in the case when all the fixed
points have a stabilizer isomorphic to $Z_2$. In this case
equations (\ref{nco2}) and (\ref{uncon}) can be satisfied by
taking  $\Gamma_{(a,i),(b,j)}$ in the  form: \be\label{gps}
\Gamma_{(a,\psi),(b,\psi')}={|G_a||G_b|\over
|G|}\hat{S}_{ab}\psi\psi'\delta_{af}\delta_{bf}\ee where $\psi$ is
the resolving index which takes two values $\pm$, and
$\hat{S}_{ab}$ is a unitary matrix. Plugging (\ref{gps}) in
(\ref{anamppp}) for the the second term one can write:
\be\label{secfo} {1\over |{\cal S}_a||{\cal
S}_d|}\psi\psi''\sum_{{\rm Orbits}\; Q(c)=0} \sum_b\sum_{J\in G}
{\hat{S}_{Ja,b}S_{c,b}\hat{S}_{b,d}^*\over
S_{0,b}}\left(\sum_{K\in
G_c}\chi_{Kc}\right)\delta_{af}\delta_{bf}\delta_{df}\ee

We also show in appendix D that formulae (\ref{anag}) and (\ref{anamppp}) are actually equivalent to the
formulae for the A-type annulus partition functions derived in \cite{Recknagel:1997sb} and \cite{Brunner:2000nk}.

\section{Gepner models: generalities}\label{genprop}
Let us remind the basic facts about Gepner models \cite{Gepner:1989gr}.
The starting point of a Gepner model is the tensor product theory
\be\label{prg}
{\cal C}^{\rm s-t}_{k_1,\cdots,k_n}=
{\cal C}^{\rm s-t}\otimes {\cal C}_{k_1}\otimes\cdots  \otimes{\cal C}_{k_n},
\ee
where ${\cal C}^{\rm s-t}$ is the $D$ dimensional flat space-time part, and ${\cal C}_{k}$ is one of the $N=2$ minimal models, whose
central charges $c_k=\frac{3k}{k+2}$ satisfy the relation
\be
\sum_{i=1}^n c_{k_i}+\frac{3}{2}(D-2)=12
\ee

$N=2$ minimal models can be described as cosets $SU(2)_k\times
U(1)_4/U(1)_{2k+4}$. Accordingly the primaries of ${\cal C}_{k}$
are labelled by three integers  $(l,m,s)$ with ranges  $l\in
(0,\cdots k)$ , $m\in (-k-1,\cdots,k+2)$, $s\in( -1,0,1,2)$,
subject to the selection rule $l+m+s\in 2Z$ and the field
identification $(l,m,s)\equiv (k-l,m+k+2, s+2)$. Primaries with
even values of $s$ belong to the NS sector, while primaries with
odd values of $s$ belong to the R sector. The conformal dimension
and charge of the primary $(l,m,s)$ are given by: \be\label{cd}
h^l_{m,s}=\frac{l(l+2)-m^2}{4(k+2)}+\frac{s^2}{8}\;\;\; ({\rm
mod}\;\; 1) \ee \be\label{ch}
q^l_{m,s}=\frac{m}{k+2}-\frac{s}{2}\;\;\; ({\rm mod}\;\; 2) \ee
The exact dimensions and charges can be read off (\ref{cd}) and
(\ref{ch}) using field identifications to bring $(l,m,s)$ into the
standard range \be l\in (0,\cdots k),\;\;\; |m-s|\leq l,\;\;\;
s\in( -1,0,1,2) \ee The characters are given by \be \label{char}
\chi^{l\,(k)}_{m,\,s}(z,\tau)=\sum_{j=0}^{k-1}c^{l\,(k)}_{m+4j-s}(\tau)\Theta_{2m+(4j-s)(k+2),2k(k+2)}(z,\tau)
\ee where \be\label{bigt}
\Theta_{M,N}(z,\tau)=\theta\left[\matrix{\frac{M}{2N}\cr
0\cr}\right](z,2N\tau)=\sum_{n\in Z}e^{2\pi i\tau
N\left(n+\frac{M}{2N}\right)^2}e^{2i\pi
z\left(n+\frac{M}{2N}\right)} \ee that obviously satisfy the
identity \be
 \Theta_{M+2N,N}=\Theta_{M,N}\ee

and $c^{l\, (k)}_m$ are the characters of the parafermionic field theory at level $k$, satisfying identities:
\be\label{iden}
 c^{l\, (k)}_m=c^{l\, (k)}_{-m}=c^{l\, (k)}_{m+2k}=c^{k-l\, (k)}_{k\pm m}
\ee
The fusion coefficients are
\bea\label{fusco}&&{\cal N}^{N=2\;\; l_1l_2l_3}_{m_1m_2m_3s_1s_2s_3}=\\ \nonumber
&&{\cal N}^{SU(2)\;l_3}_{l_1l_2}\delta_{m_1+m_2-m_3}\delta_{s_1+s_2-s_3}+{\cal N}^{SU(2)\;k-l_3}_{l_1l_2}\delta_{m_1+m_2-(m_3+k+2)}\delta_{s_1+s_2-(s_3+2)}
\eea
The space-time part can be described in terms of the $SO(D-2)_1$ algebra. $SO(2n)_1$ algebras have four primaries
$\lambda=(o,v,s,c)$, with conformal dimensions
\be
h_{o}=0,\;\;\; h_{v}=\frac{1}{2},\;\;\; h_s=h_c=\frac{n}{8}
\ee
charges
\be
q_{o}=0,\;\;\; q_{v}=1,\;\;\; q_s=\frac{n}{2}, \;\;\; q_c=\frac{n}{2}-1
\ee
and characters:
\bea\label{chso2}
&&\chi_O^{SO(2n)}=\frac{1}{2\eta^n}(\theta_3^{n}+\theta_4^{n})\\
\nonumber
&&\chi_V^{SO(2n)}=\frac{1}{2\eta^n}(\theta_3^{n}-\theta_4^{n})\\
\nonumber
&&\chi_s^{SO(2n)}=\frac{1}{2\eta^n}(\theta_2^{n}+i^{-n}\theta_1^{n})\\
\nonumber
&&\chi_c^{SO(2n)}=\frac{1}{2\eta^n}(\theta_2^{n}-i^{-n}\theta_1^{n})
\eea
$O$ and $V$ primaries belong to the NS sector, while $S$ and $C$ belong to the R sector.
For future use, let us write down also the fusion rules of the $SO(2n)_1$ algebras.
\vspace{1cm}
\bea
\begin{tabular}{||c|c|c|c|c||} \hline
n odd & o & v&s&c \\ \hline
o & o & v&s&c \\ \hline
v & v & o&c&s\\ \hline
s & s & c&v&o \\ \hline
c & c &s&o&v \\  \hline
\end{tabular}
\eea
\vspace{1cm}
\bea\label{fus}
\begin{tabular}{||c|c|c|c|c||} \hline
n even & o & v&s&c \\ \hline
o & o & v&s&c \\ \hline
v & v & o&c&s\\ \hline
s & s & c&o&v \\ \hline
c & c &s&v&o \\  \hline
\end{tabular}
\eea \vspace{1cm}

The primaries of the product theory (\ref{prg}) can be labelled by
the following collection of indices \be
(\lambda,\vec{l},\vec{m},\vec{s}))=(\lambda,l_1,m_1,s_1,\cdots,l_n,m_n,s_n)
\ee The Gepner model is the simple current extension of the
product ${\cal C}^{\rm s-t}_{k_1,\cdots,k_n}$, with the following
simple currents:
\begin{itemize}
\item  supersymmetry current: $S_{\rm tot}=(s,(0,1,1),\cdots (0,1,1))$

\item alignment currents: $V_i=(v,\cdots (0,0,2)\cdots)$, with $(0,0,2)$ at the $i$th position.

\end{itemize}

Let us summarize the results of applying the formalism reviewed in
the previous section to Gepner models \cite{Blumenhagen:1995tt},
\cite{Brunner:2000nk},\cite{Fuchs:2000gv}.
 In Gepner models the simple current projection or, in the original Gepner's language, $\beta$-projection
with respect to the supersymmetry current $S_{\rm tot}$
reads
\be\label{chharg}
Q_{(\omega,\vec{l},\vec{m},\vec{s}))}=q_{\omega}+\sum_{i=1}^n q^{l_i}_{m_i,s_i}=1 \;\; ({\rm mod}\;\;\; 2Z)
\ee
 and is nothing else than the famous GSO projection
yielding space-time supersymmetry\footnote{Actually direct
application of the formula (\ref{neut}) brings to shift $1$ with
respect to (\ref{chharg}), but as explained in
\cite{Brunner:2000nk} and \cite{Fuchs:2000gv} the shift is
absorbed by the superghost part, or alternatively by the bosonic
string map.}. The projection with respect to the alignment current
selects only primaries were all constituent primaries belong to
the same sector, either NS either R and guarantees world-sheet
supersymmetry.

To analyze the length of the orbits we should consider two cases:
\begin{enumerate}

\item all the levels $k_i$ are odd

In this case no fixed point occurs, all the primaries have trivial
stabilizer, and the length of the $S_{\rm tot}$ current is $K={\rm
lcm}\{4,2k_i+4\}$. All $V_i$ currents always act freely and have
length $2$. But when all the $k_i$ levels are odd, it turns out
that the $S_{\rm tot}$ current has an overlap with the $V_i$
currents, and to cover all orbit it is enough  to sum over only
$n-1$ of the $n$ $V_i$ currents. As a  result, the orbit length in
this case is $2^{n-1}K$.

\item one has $r\neq 0$ even levels $k_i$
\end{enumerate}

Let us place the even levels in the first $r$  positions. In this
case for a generic primary the orbit length of the supersymmetry
current is again $K={\rm lcm}\{4,2k_i+4\}$. But for the primaries
with all $l_i$ at the first $r$ positions equal $\frac{k_i}{2}$:
\be\label{lev} l_i=\frac{k_i}{2}\;\;\;\;i=1,\ldots,r \ee due to
the previously discussed field identification, which for them
reads : \bea
&&(\frac{k_1}{2},m_1,s_1,\cdots,\frac{k_r}{2},m_r,s_r,l_{r+1},m_{r+1},s_{r+1}\cdots,l_n,m_n,s_n)\equiv\\
\nonumber
&&(\frac{k_1}{2},m_1+k_1+2,s_1+2,\cdots,\frac{k_r}{2},m_r+k_r+2,s_r+2,l_{r+1},m_{r+1},s_{r+1}\cdots,l_n,m_n,s_n)
\eea there is a non-trivial stabilizer: \be {\cal
S}_{\vec{m},\vec{s}}^{\frac{k_1}{2},\cdots
\frac{k_r}{2},l_{r+1},\cdots,l_n}=Z_2. \ee We see that the
stabilizer depends only on the values of $l_i$'s $i=1,\ldots,r$
and one can write: \be |{\cal S}^{l_1,\cdots
l_r}|=1+\delta_{l_1\frac{k_1}{2}}\cdots\delta_{l_r\frac{k_r}{2}}
\ee Therefore here we have two kinds of orbits, long orbits with
length $2^nK$ for generic primary, and short orbits with length
$2^{n-1}K$ for primaries of type (\ref{lev}). As we explained the
short orbits should be resolved and acquire an additional label
$\psi$ taking two values, which we choose to be a sign $\psi=
\pm$.

\section{ The (2,2,2,2) Gepner model }\label{prim}

From now on we will specialize to the case of the $(2,2,2,2)$
Gepner model, that corresponds to a compactification down to six
dimensions. The flat part is described by an $SO(4)_1$ algebra. In
order to write down the characters of the model, first of all we
note that using the fusion rules (\ref{fus}) one can check that
the subgroup generated by the currents $S^2_{\rm tot}$ and
$V_iV_j$ has trivial action on the space-time part. The length of
the $S^2_{\rm tot}$ current is $\frac{K}{2}=4$. Using
(\ref{chharg}) we find it convenient to choose the primaries in
the form $\{v,(l_1,m_1,s_)\cdots,(l_n,m_n,s_n)\}$, with prescribed
 space-time part $v$, and neutral internal part, i.e. \be \sum_{i=1}^4
q^{l_i}_{m_i,s_i}=0\;\; ({\rm mod}\;\;\; 2Z) \ee

Now one can express the Gepner extension characters
$\chi^{G\;\vec{l}}_{(\vec{m},\vec{s})}$ in the form
\bea\label{chgep}
\chi^{G\;\vec{l}}_{(\vec{m},\vec{s})}=\frac{1}{|{\cal
S}^{\vec{l}}_{\vec{m},\vec{s}}|}\left({\cal X}_v- {\cal X}_c+{\cal
X}_o-{\cal X}_s\right) \eea where \bea\label{chgepvcos} &&{\cal
X}_v=\frac{\chi_v^{SO(4)}}{\eta^4}A(m_1,s_1,m_2,s_2,m_3,s_3,m_4,s_4)\\
\nonumber &&{\cal
X}_c=\frac{\chi_c^{SO(4)}}{\eta^4}A(m_1+1,s_1+1,m_2+1,s_2+1,m_3+1,s_3+1,m_4+1,s_4+1)
\\ \nonumber &&{\cal
X}_o=\frac{\chi_o^{SO(4)}}{\eta^4}A(m_1,s_1+2,m_2,s_2,m_3,s_3,m_4,s_4)
\\ \nonumber &&{\cal
X}_s=\frac{\chi_s^{SO(4)}}{\eta^4}A(m_1+1,s_1+3,m_2+1,s_2+1,m_3+1,s_3+1,m_4+1,s_4+1)\\
\nonumber \eea with \bea\label{geppa}
&&A(m_i,s_i)=\sum_{\nu_0=0}^3\sum_{\nu_1=0,2}\sum_{\nu_2=0,2}\sum_{\nu_3=0,2}
\chi^{l_1(2)}_{m_1+2\nu_0,\,s_1+\nu_1+\nu_2+\nu_3+2\nu_0}(z_1)\cdot\\
\nonumber
&&\chi^{l_2(2)}_{m_2+2\nu_0,\,s_2+\nu_1+2\nu_0}(z_2)\cdot
\chi^{l_3(2)}_{m_3+2\nu_0,\,s_3+\nu_2+2\nu_0}(z_3)\cdot
\chi^{l_4(2)}_{m_4+2\nu_0,\,s_4+\nu_3+2\nu_0}(z_4) \eea and, as
explained above, \be |{\cal
S}^{\vec{l}}_{\vec{m},\vec{s}}|=1+\delta_{l_11}\delta_{l_21}\delta_{l_31}\delta_{l_41}
\ee

Using (\ref{char}), (\ref{iden}) and (\ref{thetsum}) for the characters of the $k=2$ minimal model
one obtains the following simple expression

\be\label{chaar1} \chi^{l(2)}_{m,\,s}(z)=
c^{l(2)}_{m-s}(\tau)\Theta_{4q,4}(\frac{z}{2},\tau)
\ee
where $q=\frac{m}{4}-\frac{s}{2}$, and $c^{l\, (2)}_{m}$ are related to the Ising characters:
\be\label{c}
c^{0(2)}_0=\frac{1}{2\eta}\left(\sqrt{\frac{\theta_3}{\eta}}+\sqrt{\frac{\theta_4}{\eta}}\right)
\ee \be\label{cc}
c^{2(2)}_0=c^{0(2)}_2=\frac{1}{2\eta}\left(\sqrt{\frac{\theta_3}{\eta}}-\sqrt{\frac{\theta_4}{\eta}}\right)
\ee \be\label{ccc}
c^{1(2)}_1=\frac{1}{\eta}\sqrt{\frac{\theta_2}{2\eta}} \ee

Now let us compute $A(m_i,s_i)$. Repeatedly using theta functions
product formulae from appendix A, we have

\be\label{athet} A(m_i,s_i)=\Theta_{q_{\rm
tot},1}\left(\frac{z_{\rm tot}}{8},\tau\right)B(m_i,s_i) \ee where
\be
z_{\rm tot}=z_1+z_2+z_3+z_4
\ee
\be
q_{\rm tot}=q_1+q_2+q_3+q_4={\rm even}
\ee
and

\bea\label{brrr} && B(m_i,s_i)=\\ \nonumber
&&\sum_{\nu_1,\nu_2,\nu_3=0,2}
c^{l_1(2)}_{m_1-(s_1+\nu_1+\nu_2+\nu_3)}c^{l_2(2)}_{m_2-(s_2+\nu_1)}c^{l_3(2)}_{m_3-(s_3+\nu_2)}c^{l_4(2)}_{m_4-(s_4+\nu_3)}\cdot\\
\nonumber
&&\sum_{a=0,2}\Theta_{(q_1-q_2+q_3-q_4)-\nu_2+a,2}(y_1,2\tau)\cdot\\
\nonumber
&&\Theta_{(q_1-q_2-q_3+q_4)-\nu_3+a,2}(y_2,2\tau)\cdot\Theta_{(q_1+q_2-q_3-q_4)-\nu_1+a,2}(y_3,2\tau)
\eea where \bea y_1=\frac{z_1-z_2+z_3-z_4}{4},\;\;\;\;
y_2=\frac{z_1-z_2-z_3+z_4}{4},\;\;\;\;
y_3=\frac{z_1+z_2-z_3-z_4}{4} \eea Note that
$B(m_i,s_i)=B(m_i+1,s_i+1)$. Using (\ref{athet}), (\ref{chso2}),
(\ref{double}), (\ref{thetd}), this allows us to write for
(\ref{chgep}): \bea\label{chchc}
&&\chi^{G\;\vec{l}}_{(\vec{m},\vec{s})}=\\ \nonumber
&&\frac{1}{\eta^6|{\cal S}^{\vec{l}}_{\vec{m},\vec{s}}|}
\bigg[\left(\theta_2^2(2\tau)\theta_3\left(\frac{z_{\rm
tot}}{8},2\tau\right)-\theta_2(2\tau)\theta_3(2\tau)
\theta_2\left(\frac{z_{\rm tot}}{8},2\tau\right)\right)
B(m_i,s_i)\\ \nonumber
&&+\left(\theta_3^2(2\tau)\theta_2\left(\frac{z_{\rm
tot}}{8},2\tau\right)-
\theta_2(2\tau)\theta_3(2\tau)\theta_3\left(\frac{z_{\rm
tot}}{8},2\tau\right)\right) B(m_i,s_1+2,s_i))\bigg] \eea We see
that whenever \be\label{ztot} z_1+z_2+z_3+z_4=0 \ee the Gepner
extension characters are supersymmetric. This plays a role in the
study of coisotropic/magnetized D-branes and in the computations
of threshold connections \cite{pmgy1}, \cite{pmgy2}.

From now on we put all $z_i=0$. For this case the character
(\ref{chchc}) can be equivalently written as \be
\chi^{G\;\vec{l}}_{(\vec{m},\vec{s})}=\frac{\cal J}{2|{\cal
S}^{\vec{l}}_{\vec{m},\vec{s}}|\eta^6}\left(\frac{B(m_i,s_i)}{\theta_3(0,2\tau)}+
\frac{B(m_i,s_1+2,s_i)}{\theta_2(0,2\tau)}\right) \ee where ${\cal
J}=\frac{1}{2}(\theta_3^4(0,\tau)-\theta_4^4(0,\tau)-\theta_2^4(0,\tau))$
is zero thanks to Jacobi {\it aequatio identica satis abstrusa}.
Using (\ref{brrr}) and taking into account that \be\label{so2}
\Theta_{\nu,2}(z,\tau)=\eta\chi^{\rm
SO(2)}_{\nu}(\frac{z}{2},\tau) \ee as well as (\ref{c}),
(\ref{cc}), (\ref{ccc}), we are now in a position to compute the
characters for the various orbits.

To this end, we are going to present all the primaries of the
model, or in other words to list all the orbit representatives.
Surely one can pick up orbit representatives in many different
ways. To be sure that we have not taken two primaries, belonging
to the same orbit, one can resort to some kind of ``gauge fixing".
The gauge fixing chosen here, is the following.
\begin{enumerate}

\item We take the space-time part to be always $v$, as mentioned
above.

\item we take $s_2=s_3=s_4=0$

\item we limit $m_1$ to the values 0 and 1.

\item to avoid taking primaries equivalent due to field
identification, we always limit the values of the $l_i$ to be 0 or
1.

\end{enumerate}

The final picture is the following.

In this model we can divide primaries in 4 big groups.

The first group has $l_1=l_2=l_3=l_4=0$, $s_1=s_2=s_3=s_4=0$ and
contains 16 primaries. We can divide them into three groups:
$K_1$, $K_2$ and $K_3$. All primaries in the same group have the
same conformal weights and characters. The results are presented
in the tables below. It is understood that all the entries  should
be multiplied by $\frac{\cal J}{\eta^{12}}$.

 \vspace{1cm}
\be\label{chK1}
 \begin{tabular}{||c|c||} \hline
 $K_1$&$h_{K_1}=\frac{1}{2}$\\ \hline
$K_1=(v)(0,0,0)(0,0,0)(0,0,0)(0,0,0)$& $\chi_{K_{1}}^G=\frac{\theta_3^4(0,\tau)+\theta_4^4(0,\tau)}{16}+
3\frac{\theta_3^2(0,\tau)\theta_4^2(0,\tau)}{8}$\\ \hline
\end{tabular}
\ee
\vspace{1cm}
\be
 \begin{tabular}{||c|c||} \hline
$K_2$& $h_{K_2}=1$\\ \hline
$K_{2a}=(v)(0,0,0)(0,0,0)(0,-2,0)(0,2,0)$&\\
$K_{2b}=(v)(0,0,0)(0,-2,0)(0,0,0)(0,2,0)$&\\
$K_{2c}=(v)(0,0,0)(0,2,0)(0,-2,0)(0,0,0)$&\\
$K_{2d}=(v)(0,0,0)(0,0,0)(0,2,0)(0,-2,0)$&\\
$K_{2e}=(v)(0,0,0)(0,-2,0)(0,2,0)(0,0,0)$&\\
$K_{2f}=(v)(0,0,0)(0,2,0)(0,0,0)(0,-2,0)$&$\chi_{K_{2}}^G=\frac{\theta_3^4(0,\tau)-\theta_4^4(0,\tau)}{16}$\\
$K_{2g}=(v)(0,0,0)(0,4,0)(0,2,0)(0,2,0)$&\\
$K_{2h}=(v)(0,0,0)(0,2,0)(0,4,0)(0,2,0)$&\\
$K_{2k}=(v)(0,0,0)(0,2,0)(0,2,0)(0,4,0)$&\\
$K_{2l}=(v)(0,0,0)(0,4,0)(0,-2,0)(0,-2,0)$&\\
$K_{2m}=(v)(0,0,0)(0,-2,0)(0,4,0)(0,-2,0)$&\\
$K_{2n}=(v)(0,0,0)(0,-2,0)(0,-2,0)(0,4,0)$&\\ \hline
 \end{tabular}
 \ee

\vspace{1cm}

   \be
   \begin{tabular}{||c|c||} \hline
   $K_{3}$&$h_{K_3}=\frac{3}{2}$\\ \hline
 $K_{3a}=(v)(0,0,0)(0,0,0)(0,4,0)(0,4,0)$&\\
$K_{3b}=(v)(0,0,0)(0,4,0)(0,0,0)(0,4,0)$&$\chi_{K_{3}}^G=\frac{\theta_3^4(0,\tau)+\theta_4^4(0,\tau)}{16}-
\frac{\theta_3^2(0,\tau)\theta_4^2(0,\tau)}{8}$\\
$K_{3c}=(v)(0,0,0)(0,4,0)(0,4,0)(0,0,0)$&\\ \hline
\end{tabular}
 \ee

\vspace{1cm}

The second group has $l_1=l_2=l_3=l_4=0$, $s_1=2$,
$s_2=s_3=s_4=0$ and also contains  16 primaries,
 which again can be divided into 3 subgroups, in such a way that all primaries inside each group
 have the same characters.

 \be
  \begin{tabular}{||c|c||} \hline
  $L_1$& $h_{L_1}=1$\\ \hline
$L_{1a}=(v)(0,0,2)(0,4,0)(0,4,0)(0,4,0)$&\\
 $L_{1b}=(v)(0,0,2)(0,0,0)(0,0,0)(0,4,0)$&\\
$L_{1c}=(v)(0,0,2)(0,0,0)(0,4,0)(0,0,0)$&$\chi_{L_{1}}^G=
\frac{\theta_3^4(0,\tau)-\theta_4^4(0,\tau)}{16}$\\
$L_{1d}=(v)(0,0,2)(0,4,0)(0,0,0)(0,0,0)$&\\ \hline
\end{tabular}
 \ee

\vspace{1cm}

  \be
 \begin{tabular}{||c|c||} \hline
  $L_2$&$h_{L_2}=\frac{1}{2}$\\ \hline
$L_{2a}=(v)(0,0,2)(0,4,0)(0,2,0)(0,-2,0)$&\\
$L_{2b}=(v)(0,0,2)(0,-2,0)(0,2,0)(0,4,0)$&\\
$L_{2c}=(v)(0,0,2)(0,-2,0)(0,4,0)(0,2,0)$&$\chi_{L_{2}}^G=
\frac{\theta_3^4(0,\tau)+\theta_4^4(0,\tau)}{16}+
\frac{\theta_3^2(0,\tau)\theta_4^2(0,\tau)}{8}$\\
$L_{2d}=(v)(0,0,2)(0,4,0)(0,-2,0)(0,2,0)$&\\
$L_{2e}=(v)(0,0,2)(0,2,0)(0,-2,0)(0,4,0)$&\\
$L_{2f}=(v)(0,0,2)(0,2,0)(0,4,0)(0,-2,0)$&\\ \hline
\end{tabular}
 \ee

 \vspace{1cm}

   \be
\begin{tabular}{||c|c||} \hline
 $L_{3}$& $h_{L_3}=\frac{3}{2}$\\ \hline
$L_{3a}=(v)(0,0,2)(0,0,0)(0,2,0)(0,2,0)$&\\
$L_{3b}=(v)(0,0,2)(0,2,0)(0,0,0)(0,2,0)$&\\
$L_{3c}=(v)(0,0,2)(0,2,0)(0,2,0)(0,0,0)$&$\chi_{L_{3}}^G=
\frac{\theta_3^4(0,\tau)+\theta_4^4(0,\tau)}{16}-
\frac{\theta_3^2(0,\tau)\theta_4^2(0,\tau)}{8}$\\
$L_{3d}=(v)(0,0,2)(0,0,0)(0,-2,0)(0,-2,0)$&\\
$L_{3e}=(v)(0,0,2)(0,-2,0)(0,-2,0)(0,0,0)$&\\
$L_{3f}=(v)(0,0,2)(0,-2,0)(0,0,0)(0,-2,0)$&\\ \hline
\end{tabular}
 \ee

 \vspace{1cm}

The third group containing 48 primaries with any two of $l_i$ equal to $1$, and other two
of them to $0$. This group consists of 6 subgroups:
\bea
&&l_1=l_2=1\quad l_3=l_4=0\\ \nonumber
&&l_1=l_3=1\quad l_2=l_4=0\\ \nonumber
&&l_1=l_4=1\quad l_2=l_3=0\\ \nonumber
&&l_2=l_3=1\quad l_1=l_4=0\\ \nonumber
&&l_2=l_4=1\quad l_1=l_3=0\\ \nonumber
&&l_3=l_4=1\quad l_1=l_2=0
\eea
Each such a subgroup consists of 8 primaries and
 can be derived from, let's say, the first of them by permutations, so we will
write down only one of them, the one with  $l_1=l_2=1$ and
$l_3=l_4=0$. We schematically denote the primaries in this group
as $\Phi_{1a}^{1,1,\cdot,\cdot}$, indicating explicitly in the
superscript which $l_i$ are equal to 1.

\be
\begin{tabular}{||c|c||} \hline
$\Phi_1$&$h_{\Phi_1}=\frac{3}{4}$\\ \hline
$\Phi_{1a}^{1,1,\cdot,\cdot}=(v)(1,1,0)(1,3,0)(0,2,0)(0,2,0)$&\\
$\Phi_{1b}^{1,1,\cdot,\cdot}=(v)(1,1,0)(1,-1,0)(0,0,0)(0,0,0)$&$\chi_{\Phi_1}^G=
\frac{\theta_2^2(0,\tau)(\theta_3^2(0,\tau)+\theta_4^2(0,\tau))}{8}$\\ \hline
\end{tabular}
\ee

 \vspace{1cm}

\be
\begin{tabular}{||c|c||} \hline
$\Phi_2$&$h_{\Phi_2}=\frac{5}{4}$\\ \hline
$\Phi_{2a}^{1,1,\cdot,\cdot}=(v)(1,1,0)(1,3,0)(0,0,0)(0,4,0)$&\\
$\Phi_{2b}^{1,1,\cdot,\cdot}=(v)(1,1,0)(1,-1,0)(0,-2,0)(0,2,0)$&$\chi_{\Phi_2}^G=
\frac{\theta_2^2(0,\tau)(\theta_3^2(0,\tau)-\theta_4^2(0,\tau))}{8}$\\ \hline
\end{tabular}
\ee

\vspace{1cm}

\be
\begin{tabular}{||c|c||} \hline
$\Phi_3$&$h_{\Phi_3}=1$\\ \hline
 $\Phi_{3a}^{1,1,\cdot,\cdot}=(v)(1,1,0)(1,-3,0)(0,2,0)(0,0,0)$&\\
$\Phi_{3b}^{1,1,\cdot,\cdot}=(v)(1,1,0)(1,-3,0)(0,0,0)(0,2,0)$&$\chi_{\Phi_3}^G=
\frac{\theta_3^4(0,\tau)-\theta_4^4(0,\tau)}{8}$\\ \hline
\end{tabular}
\ee

\vspace{1cm}

\be
\begin{tabular}{||c|c||} \hline
$\Phi_4$&$h_{\Phi_4}=\frac{1}{2}$\\ \hline
$\Phi_{4a}^{1,1,\cdot,\cdot}=(v)(1,1,0)(1,1,0)(0,-2,0)(0,0,0)$&\\
$\Phi_{4a}^{1,1,\cdot,\cdot}=(v)(1,1,0)(1,1,0)(0,0,0)(0,-2,0)$&$\chi_{\Phi_{4}}^G=
\frac{\theta_3^4(0,\tau)+\theta_4^4(0,\tau)}{8}$\\ \hline
\end{tabular}
 \ee

 \vspace{1cm}

Finally  we have a small group containing only 4 elements with
$l_1=l_2=l_3=l_4=1$, $s_1=s_2=s_3=s_4=0$. All primaries in this
group, as we explained in section \ref{genprop}, have a short
orbit and should be resolved. After resolution we end up with 8
primaries. The $\pm$ in the notations refers to the resolution
process.

   \be
\begin{tabular}{||c|c||} \hline
$R_1$&   $h_{R_1}=1$\\ \hline
$R_{1a\pm}=(v)(1,1,0)(1,-1,0)(1,1,0)(1,-1,0)_{\pm}$&\\
$R_{1b\pm}=(v)(1,1,0)(1,-1,0)(1,-1,0)(1,1,0)_{\pm}$&$\chi_{R_1}^G= \
\frac{\theta^4_2(0,\tau)}{8}$\\
$R_{1c\pm}=(v)(1,1,0)(1,1,0)(1,-1,0)(1,-1,0)_{\pm}$&\\ \hline
\end{tabular}
\ee

\vspace{1cm}

\be
\begin{tabular}{||c|c||} \hline
$R_2$& $h_{R_2}=\frac{1}{2}$\\ \hline
 $R_{2\pm}=(v)(1,1,0)(1,1,0)(1,-3,0)(1,1,0)_{\pm}$&$\chi_{R_2}^G=
\frac{\theta^4_3(0,\tau)+\theta^4_4(0,\tau)}{8}$\\ \hline
\end{tabular}
 \ee

 \vspace{1cm}

We see that before fixed points resolution we had 84 orbits: 31 orbits
with conformal dimension $1$, 12 orbits with conformal dimension
$\frac{3}{4}$, 12 orbits with conformal dimension $\frac{5}{4}$,
20 orbits with conformal dimension $\frac{1}{2}$, 9 orbits
with conformal dimension $\frac{3}{2}$. After the fixed points
resolution we have 88 primaries:
 34 orbits with conformal dimension $1$, 12 orbits with conformal dimension
$\frac{3}{4}$, 12 orbits with conformal dimension $\frac{5}{4}$,
21 orbits with conformal dimension $\frac{1}{2}$, 9 orbits
with conformal dimension $\frac{3}{2}$ \cite{yas}.

Collecting all the above results, we can write down the torus
amplitude: \bea\label{pfge} && Z=\left|\frac{{\cal
J}}{\eta^{12}}\right|^2\left[\left|\left(\frac{\theta_3^4(0,\tau)+\theta_4^4(0,\tau)}{16}+
3\frac{\theta_3^2(0,\tau)\theta_4^2(0,\tau)}{8}\right)\right|^2\right.\\
\nonumber
&&16\left|\frac{\theta_3^4(0,\tau)-\theta_4^4(0,\tau)}{16}\right|^2+9\left|\frac{\theta_3^4(0,\tau)+\theta_4^4(0,\tau)}{16}-
\frac{\theta_3^2(0,\tau)\theta_4^2(0,\tau)}{8}\right|^2\\
\nonumber
&&+6\left|\frac{\theta_3^4(0,\tau)+\theta_4^4(0,\tau)}{16}+
\frac{\theta_3^2(0,\tau)\theta_4^2(0,\tau)}{8}\right|^2+18\left|\frac{\theta_3^4(0,\tau)-\theta_4^4(0,\tau)}{8}\right|^2\\
\nonumber
&&+14\left|\frac{\theta_3^4(0,\tau)+\theta_4^4(0,\tau)}{8}\right|^2+12\left|
\frac{\theta_2^2(0,\tau)(\theta_3^2(0,\tau)+\theta_4^2(0,\tau))}{8}\right|^2+\\
\nonumber &&\left.12\left|
\frac{\theta_2^2(0,\tau)(\theta_3^2(0,\tau)-\theta_4^2(0,\tau))}{8}\right|^2\right]\\
\nonumber &&=\left|\frac{{\cal
J}}{\eta^{12}}\right|^2\left[\frac{1}{16}\bigg(|\theta_3(0,\tau)|^4+|\theta_4(0,\tau)|^4+|\theta_2(0,\tau)|^4\bigg)^2
\right.\\ \nonumber
&&+\frac{1}{4}\bigg(|\theta_2(0,\tau)\theta_3(0,\tau)|^4+|\theta_2(0,\tau)\theta_4(0,\tau)|^4+|\theta_3(0,\tau)(\theta_4(0,\tau)|^4\bigg)\\
\nonumber
&&\left.+\frac{1}{2}\bigg(|\theta_3(0,\tau)|^8+|\theta_4(0,\tau)|^8\bigg)\right]
\eea

The partition function (\ref{pfge}), as first noted in
\cite{Eguchi:1988vr}, coincides with the partition function of the
$T^4/Z_4$ orbifold at the $SU(2)^4$ point, which we review in
appendix C.

Now we elaborate on the expression (\ref{anamppp}) for the annulus partition function for the $(2,2,2,2,)$ Gepner model.

Let us denote the first Cardy state $I$: \be
I=(S_0,L_1,M_1,S_1,\cdots, L_4,M_4,S_4) \ee and the second $J$:
\be J=(\tilde{S}_0,\tilde{L}_1,\tilde{M}_1,\tilde{S}_1,\cdots,
\tilde{L}_4,\tilde{M}_4,\tilde{S}_4) \ee Consider first the case
when neither the first boundary state nor the second are fixed.

Now using (\ref{anag}) and the fusion coefficients (\ref{fusco})
we can easily derive: \be\label{p2}
Z_{IJ}=\sum_{s_0}\sum_{l_i}|{\cal S}^{l_1\cdots l_4}|{\cal
N}_{v(S_0)\tilde{S}_0}^{SO(4)\,s_0} \prod_{i=1}^4{\cal
N}_{L_i\tilde{L}_i}^{SU(2)\,l_i}\chi^{G\, \hat{l_1}\cdots
\hat{l_4}}_{\hat{M_1-\tilde{M}_1}\cdots \hat{M_4-\tilde{M}_4},
\hat{s_0},\hat{S_1-\tilde{S}_1}\cdots \hat{S_4-\tilde{S}_4}} \ee
Actually the sum over $J$ in (\ref{anag}) $\sum_{J\in G}{\cal
N}_{Ja,c}^d$ is running over the orbit of the primary
\be\label{difp} (s_0,l_1,M_1-\tilde{M}_1,S_1-\tilde{S}_1\cdots
l_4,M_4-\tilde{M}_4,S_4-\tilde{S}_4) \ee while the sum over orbits
in (\ref{anag}) runs over the specific representatives, for
examples listed in the tables above. It means that generically in
this sum only one term will survive, the specific representative
of the orbit of the primary (\ref{difp}). If this primary has
non-trivial stabilizer, due to field identification the sum over
$J$ will produce the representative twice. The fusion $v(S_0)$ in
${\cal N}_{v(S_0)\tilde{S}_0}^{SO(4)}$ is due to the bosonic
string map \cite{Fuchs:2000gv}. Collecting all pieces we get
(\ref{p2}). In practice in order to use formula (\ref{p2}) one
needs to compute the primary (\ref{difp}) and then use the action
of the simple current to find in the orbit which of the
representatives listed in tables above it belongs to, and
substitute its character.

Consider next the case when $I$ is not fixed but $J$ is. In this
case elaborating on (\ref{anaff}) we obtain: \be\label{p3}
Z_{IJ}=\frac{1}{|{\cal S}_J|}\sum_{s_0}\sum_{l_i}|{\cal
S}^{l_1\cdots l_4}|{\cal N}_{v(S_0)\tilde{S}_0}^{SO(4)\,s_0}
\prod_{i=1}^4{\cal N}_{L_i\tilde{L}_i}^{SU(2)\,l_i}\chi^{G\,
\hat{l}_1\cdots \hat{l}_4}_{\hat{M_1-\tilde{M}_1}\cdots
\hat{M_4-\tilde{M}_4}, \hat{s}_0,\hat{S_1-\tilde{S}_1}\cdots
\hat{S_4-\tilde{S}_4}} \ee

The last case is when both and $I$ and $J$ are fixed points. To
elaborate on this case we need the matrices $S_{c,b}$ and
$\hat{S}_{ab}$ in formula (\ref{secfo}).

The matrix $S_{ab}$ for Gepner models is the product of  all the
elementary $S$'s and reads: \be\label{gepm}
2^4S^{SO(4)}_{s_0s'_0}\prod_{i=1}^4
S^{SU(2)_{k_i}}_{l_il'_{i}}S^{U(1)_4}_{s_is'_{i}}S^{U(1)_{k+2}}_{m_im'_{i}}
\ee The matrix $\hat{S}_{ab}$ was found in  \cite{Fuchs:2000gv}.
For the (2,2,2,2) model it has the form \be\label{fixm}
-2^4S^{SO(4)}_{s_0s'_0}\prod_{i=1}^4
S^{U(1)_4}_{s_is'_{i}}S^{U(1)_{k+2}}_{m_im'_{i}} \ee The numerical
factors come from the field identification.

Plugging (\ref{gepm}) and (\ref{fixm}) in (\ref{secfo}) one obtains:
\bea\label{p4}
&&Z_{I\psi J\psi'}=
\frac{1}{|{\cal S}_I||{\cal S}_J|}\sum_{s_0}\sum_{l_i}{\cal N}_{v(S_0)\tilde{S}_0}^{SO(4)\,s_0}\cdot\\ \nonumber
&&\left(\prod_{i=1}^4{\cal N}_{L_i\tilde{L}_i}^{SU(2)\,l_i}+\psi\psi'
\prod_{i=1}^{4}\sin\pi\frac{l_i+1}{2}
\right)\chi^{G\, \hat{l}_1\cdots \hat{l}_4}_{\hat{M_1-\tilde{M}_1}\cdots \hat{M_4-\tilde{M}_4},
\hat{s}_0,\hat{S_1-\tilde{S}_1}\cdots \hat{S_4-\tilde{S}_4}}
\eea

\section{$D0$-branes on the $T^4/Z_4$ orbifold.}\label{brorb}

\subsection{Fixed points}

Defining complex coordinates $z_1=x_1+ix_2$ and $z_2=x_3+ix_4$ the $Z_4$ group action can be described as
\be z_1 \rightarrow e^{\frac{2i\pi k}{4}} z_1 \quad z_2 \rightarrow e^{-\frac{2i\pi k}{4}} z_2
\ee
We can consider it as generated by the $Z_2$ subgroup acting as $z_1\rightarrow -z_1$ and $z_2\rightarrow -z_2$
and a $Z'_2$ subgroup rotating by $\frac{\pi}{2}$ and $-\frac{\pi}{2}$ the $(x_1,x_2)$ and $(x_3,x_4)$ planes: $z_1\rightarrow iz_1$ and
 $z_2\rightarrow -iz_2$.

The $Z_2$ group has 16 fixed points $(\pi Re_1,\pi Re_2,\pi Re_3,\pi Re_4)$, where $e_i=0,1$, out of which
only 4 are also fixed under $Z'_2$ : $D0_{1f}=(0,0,0,0)$,  $D0_{2f}=(\pi R,\pi R,\pi R,\pi R)$, $D0_{3f}=(\pi R,\pi R,0,0)$,
$D0_{4f}=(0,0,\pi R,\pi R)$.

To begin with, let us calculate the annulus partition function for
open strings having both ends at the same fixed point.

The partition function is given by
\bea &&Z_{D0_fD0_f}=\\ \nonumber
&&\frac{1}{8}\sum_{k=0}^3{\rm Tr}(1+(-)^F)g^k e^{-2\pi\tau
L_0}=\frac{1}{4}\frac{{\cal J}}{\eta^{12}}Z_{\rm
windings}+\frac{1}{8}\sum_{k=1}^3(4\sin^2\frac{\pi k}{4})Z_{0,k}
\eea

where
 \bea\label{untwp} &&Z_{\rm
windings}=\\ \nonumber
&&\sum_{n_1,n_2,n_3,n_4}q^{n_1^2+n_2^2+n_3^2+n_4^2}=\theta_3^4(0,2\tau)=
\frac{\theta_3^4(0,\tau)+\theta_4^4(0,\tau)}{4}+\frac{\theta_3^2(0,\tau)
\theta_4^2(0,\tau)}{2} \eea

and $Z_{0,k}$ can be found in (\ref{z01}), (\ref{z02}),
(\ref{z03}) of appendix C. Collecting all the pieces, we obtain:

\be\label{pfD1}
Z_{D0_fD0_f}=\frac{{\cal J}}{\eta^{12}}\left(\frac{\theta_3^4(0,\tau)+\theta_4^4(0,\tau)}{16}+\frac{3\theta_3^2(0,\tau)
\theta_4^2(0,\tau)}{8}\right)
\ee
We see that (\ref{pfD1}) coincides with (\ref{chK1}):
\be
Z_{D0_fD0_f}=\chi_{K_1}
\ee

In order to compute the partition function for strings with ends
at different fixed point, we need to recall the partition function
for a scalar $X$ compactified at the self-dual radius
$R=\frac{1}{\sqrt{2}}$ with Dirichlet boundary conditions placed
at $2\pi R\xi_1$ and $2\pi R\xi_2$, so that
 \be X=2\pi R\xi_1+\left(2R(\xi_2-\xi_1)+2nR\right)\sigma+{\rm
oscillators} \ee
The partition function is easily calculated to be
\be\label{pfsca}
Z_{x_1x_2}=\frac{1}{\eta}q^{(\xi_2-\xi_1)^2}\theta_3(2\tau
(\xi_2-\xi_1), 2\tau) \ee

Using (\ref{pfsca}) we can then compute the annulus partition functions between different fixed points:
\be
Z_{D0_{1f}D0_{2f}}=\frac{{\cal J}}{\eta^{12}}\left(\frac{\theta_3^4(0,\tau)+\theta_4^4(0,\tau)}{16}+\frac{\theta_3^2(0,\tau)
\theta_4^2(0,\tau)}{8}\right)=\chi_{L_2}
\ee

\be\label{df3}
Z_{D0_{1f}D0_{3f}}=\frac{{\cal J}}{\eta^{12}}\left(\frac{\theta_3^4(0,\tau)-\theta_4^4(0,\tau)}{16}+\frac{\theta_3^2(0,\tau)
\theta_4^2(0,\tau)}{4}\right)
\ee

It seems that (\ref{df3}) does not fall in the list of characters
computed in section \ref{prim}. We think it means that the
$D0_{f3}$ cannot be described by a Cardy state, and do not
consider it any further here.

\subsection{ Partially fixed points}

Now we consider the case when the $D0$ branes lie at a point fixed
only under $Z_2$. We have the following list of such branes:
 \bea\label{pfbr}
&&D0_1=A_1+A_1':(0,\pi R, 0,0)+(\pi R,0, 0,0)\\ \nonumber
&&D0_2=A_2+A_2':(\pi R, 0,0,\pi R)+(0,\pi R,\pi R,0)\\ \nonumber
&&D0_3=A_3+A_3':(\pi R,\pi R, 0,\pi R)+(\pi R,\pi R,\pi R,0)\\ \nonumber
&&D0_4=A_4+A_4':(\pi R,0,\pi R,\pi R)+(0,\pi R,\pi R,\pi R)\\\nonumber
&&D0_5=A_5+A_5':(0,0, 0,\pi R)+(0,0,\pi R,0)\\ \nonumber
&&D0_6=A_6+A_6':(\pi R,0,\pi R,0)+(0,\pi R,0,\pi R)
 \eea

The partition functions between branes (\ref{pfbr}) and fixed point branes are given by equation:
\be\label{parimm2}
Z_{D0_{i}D0_{f}}={\rm Tr}_{A_iD0_f}\frac{(1+(-)^F)}{2}\frac{(1+g^2)}{2}e^{-2\pi\tau L_0}
\ee
which taking into account (\ref{z02}) simplifies to
\be\label{parim}
Z_{D0_{i}D0_{f}}={\rm Tr}_{A_iD0_f}\frac{(1+(-)^F)}{4}e^{-2\pi\tau L_0}
\ee

Using (\ref{pfsca}) we can easily compute all annulus partition
functions of this type. The result is presented in the following
table: \be\label{tabp}
\begin{tabular}{||c|c|c||} \hline
Branes& $D0_{1f}$& $D0_{2f}$\\ \hline
$D0_{1f}$&$\frac{\theta_3^4+\theta_4^4}{16}+\frac{3\theta_3^2\theta_4^2}{8}$
&$\frac{\theta_3^4+\theta_4^4}{16}+\frac{\theta_3^2\theta_4^2}{8}$\\
\hline
$D0_{2f}$&$\frac{\theta_3^4+\theta_4^4}{16}+\frac{\theta_3^2\theta_4^2}{8}$
&$\frac{\theta_3^4+\theta_4^4}{16}+\frac{3\theta_3^2\theta_4^2}{8}$\\
\hline
$D0_1$&$\frac{\theta_2^2(\theta_3^2+\theta_4^2)}{8}$&$\frac{\theta_2^2(\theta_3^2-\theta_4^2)}{8}$\\
\hline
$D0_2$&$\frac{\theta_3^4-\theta_4^4}{8}$&$\frac{\theta_3^4-\theta_4^4}{8}$\\
\hline $D0_3$&$\frac{\theta_2^2(\theta_3^2-\theta_4^2)}{8}$
&$\frac{\theta_2^2(\theta_3^2+\theta_4^2)}{8}$\\ \hline
$D0_4$&$\frac{\theta_2^2(\theta_3^2-\theta_4^2)}{8}$&$\frac{\theta_2^2(\theta_3^2+\theta_4^2)}{8}$\\
\hline $D0_5$& $\frac{\theta_2^2(\theta_3^2+\theta_4^2)}{8}$&
$\frac{\theta_2^2(\theta_3^2-\theta_4^2)}{8}$\\ \hline
$D0_6$&$\frac{\theta_3^4-\theta_4^4}{8}$&$\frac{\theta_3^4-\theta_4^4}{8}$\\
\hline
\end{tabular}
\ee \vspace{1cm}

where it is understood that all entries should be multiplied by
$\frac{{\cal
J}}{\eta^{12}}=\frac{1}{2\eta^{12}}(\theta_3^4-\theta_4^4-\theta_2^4)$.

Using the characters in section \ref{prim} one can present table
(\ref{tabp}) in the form

\be\label{pfkar}
\begin{tabular}{||c|c|c||} \hline
Branes& $D0_{1f}$& $D0_{2f}$\\ \hline $D0_{1f}$&$\chi_{K_1}$
&$\chi_{L_2}$\\
\hline $D0_{2f}$&$\chi_{L_2}$
&$\chi_{K_1}$\\
\hline
$D0_1$&$\chi_{\Phi_1}$&$\chi_{\Phi_2}$\\
\hline
$D0_2$&$\chi_{R_1}$&$\chi_{R_1}$\\
\hline $D0_3$&$\chi_{\Phi_2}$ &$\chi_{\Phi_1}$\\
\hline
$D0_4$&$\chi_{\Phi_2}$&$\chi_{\Phi_1}$\\
\hline $D0_5$& $\chi_{\Phi_1}$& $\chi_{\Phi_2}$\\
\hline
$D0_6$&$\chi_{R_1}$&$\chi_{R_1}$\\
\hline
\end{tabular}
\ee

\vspace{1cm}

Table (\ref{pfkar}) already gives a hint for the candidate Cardy
states, describing $D0$ branes located at fixed and partially
fixed points.

To make things more precise we should compute also the partition
functions between the different partially fixed branes
(\ref{pfbr}). They have the form: \be\label{parim3}
Z_{D0_{i}D0_{j}}={\rm
Tr}_{A_iA_j}\frac{(1+(-)^F)}{2}\frac{(1+g^2)}{2}e^{-2\pi\tau L_0}+
{\rm Tr}_{A_iA_j'}\frac{(1+(-)^F)}{2}\frac{(1+g^2)}{2}e^{-2\pi\tau
L_0} \ee which using (\ref{z02}) simplifies to \be\label{parimm}
Z_{D0_{i}D0_{j}}={\rm Tr}_{A_iA_j}\frac{(1+(-)^F)}{4}e^{-2\pi\tau
L_0}+ {\rm Tr}_{A_iA_j'}\frac{(1+(-)^F)}{4}e^{-2\pi\tau L_0} \ee

\vspace{1cm}

Again using (\ref{pfsca}) we can present the result in the following table:

\vspace{1cm}
\be\label{dtta}
\begin{tabular}{||c|c|c|c|c|c|c||} \hline
Branes   & $D0_1$ &$D0_2$& $D0_3$&$D0_4$ &$D0_5$&$D0_6$ \\
\hline $D0_1$ &$\frac{\theta_3^4}{4}+\frac{\theta_3^2\theta_4^2}{4}$
&$\frac{\theta_2^2\theta_3^2}{4}$ &$\frac{\theta_3^4-\theta_4^4}{4}$
& $\frac{\theta_3^4}{4}-\frac{\theta_3^2\theta_4^2}{4}$
&$\frac{\theta_3^4-\theta_4^4}{4}$& $\frac{\theta_2^2\theta_3^2}{4}$\\
\hline $D0_2$&
$\frac{\theta_2^2\theta_3^2}{4}$&$\frac{\theta_3^4+\theta_4^4}{4}$ &
$\frac{\theta_2^2\theta_3^2}{4}$& $\frac{\theta_2^2\theta_3^2}{4}$
&$\frac{\theta_2^2\theta_3^2}{4}$& $\frac{\theta_3^4-\theta_4^4}{4}$\\
\hline $D0_3$& $\frac{\theta_3^4-\theta_4^4}{4}$&
$\frac{\theta_2^2\theta_3^2}{4}$&
  $\frac{\theta_3^4}{4}+\frac{\theta_3^2\theta_4^2}{4}$ &$\frac{\theta_3^4-\theta_4^4}{4}$
&$\frac{\theta_3^4}{4}-\frac{\theta_3^3\theta_4^2}{4}$&$\frac{\theta_2^2\theta_3^2}{4}$
\\ \hline $D0_4$&
$\frac{\theta_3^4}{4}-\frac{\theta_3^2\theta_4^2}{4}$
 &$\frac{\theta_2^2\theta_3^2}{4}$ &$\frac{\theta_3^4-\theta_4^4}{4}$ & $\frac{\theta_3^4}{4}+\frac{\theta_3^2\theta_4^2}{4}$
&$\frac{\theta_3^4-\theta_4^4}{4}$& $\frac{\theta_2^2\theta_3^2}{4}$\\
\hline $D0_5$& $\frac{\theta_3^4-\theta_4^4}{4}$&
$\frac{\theta_2^2\theta_3^2}{4}$&$\frac{\theta_3^4}{4}-\frac{\theta_3^3\theta_4^2}{4}$&$\frac{\theta_3^4-\theta_4^4}{4}$
&$\frac{\theta_3^4}{4}+\frac{\theta_3^2\theta_4^2}{4}$&$\frac{\theta_2^2\theta_3^2}{4}$
\\ \hline $D0_6$
&$\frac{\theta_2^2\theta_3^2}{4}$&$\frac{\theta_3^4-\theta_4^4}{4}$&
$\frac{\theta_2^2\theta_3^2}{4}$&$\frac{\theta_2^2\theta_3^2}{4}$&$\frac{\theta_2^2\theta_3^2}{4}$
& $\frac{\theta_3^4+\theta_4^4}{4}$\\ \hline
\end{tabular}
\ee

\vspace{1cm}

where, as before, it is understood that all entries should be
multiplied by $\frac{{\cal
J}}{\eta^{12}}=\frac{1}{2\eta^{12}}(\theta_3^4-\theta_4^4-\theta_2^4)$.

After some trial and error we can solve these conditions with the following
Cardy states: \bea\label{cs} && D0_{1f}=|K_1\rangle^{\rm Cardy}\\ \nonumber &&
D0_{2f}=|L_{2a}\rangle^{\rm Cardy}\\ \nonumber &&
D0_1=|\Phi_{1a}^{1,1,\cdot,\cdot}\rangle^{\rm Cardy}\\ \nonumber &&
D0_2=|R_{1a+}\rangle^{\rm Cardy}\\ \nonumber &&
D0_3=|\Phi_{2b}^{\cdot,\cdot,1,1}\rangle^{\rm Cardy}\\ \nonumber &&
D0_4=|\Phi_{2b}^{1,1,\cdot,\cdot}\rangle^{\rm Cardy}\\ \nonumber &&
D0_5=|\Phi_{1a}^{\cdot,\cdot,1,1}\rangle^{\rm Cardy}\\ \nonumber &&
D0_6=|R_{1a-}\rangle^{\rm Cardy}\\ \nonumber \eea

Using the formulae (\ref{p2}),(\ref{p3}),(\ref{p4}) we obtain for the annulus partition functions between the
states (\ref{cs}) the following table :

\vspace{1cm}

\begin{tabular}{||c|c|c|c|c|c||} \hline
Branes   & $D0_1$ &$D0_2$& $D0_3$&$D0_4$ &$D0_5$\\
\hline $D0_1$ &$Z_{11}$
&$\chi_{\Phi_2}+\chi_{\Phi_1}$ &$2\chi_{R_1}$ & $2(\chi_{K_3}+\chi_{L_1})$
&$2\chi_{R_1}$\\
\hline $D0_2$& $\chi_{\Phi_2}+\chi_{\Phi_1}$&$\chi_{K_1}+3\chi_{K_3}$ &
$\chi_{\Phi_2}+\chi_{\Phi_1}$& $\chi_{\Phi_1}+\chi_{\Phi_2}$
&$\chi_{\Phi_1}+\chi_{\Phi_2}$\\
\hline $D0_3$& $2\chi_{R_1}$&
$\chi_{\Phi_2}+\chi_{\Phi_1}$&
  $Z_{11}$ &$2\chi_{R_1}$
&$2(\chi_{K_3}+\chi_{L_1})$
\\ \hline $D0_4$&
$2(\chi_{K_3}+\chi_{L_1})$
 &$\chi_{\Phi_1}+\chi_{\Phi_2}$ &$2\chi_{R_1}$ & $Z_{11}$
&$2\chi_{R_1}$\\
\hline $D0_5$& $2\chi_{R_1}$&
$\chi_{\Phi_1}+\chi_{\Phi_2}$&$2(\chi_{K_3}+\chi_{L_1})$&$2\chi_{R_1}$
&$Z_{11}$
\\ \hline $D0_6$
&$\chi_{\Phi_2}+\chi_{\Phi_1}$&$4\chi_{L_1}$&
$\chi_{\Phi_2}+\chi_{\Phi_1}$& $\chi_{\Phi_1}+\chi_{\Phi_2}$
&$\chi_{\Phi_1}+\chi_{\Phi_2}$
\\ \hline
\end{tabular}

\vspace{1cm}

where $Z_{11}=\chi_{K_1}+\chi_{K_3}+2\chi_{L_1}$, which coincides with table (\ref{dtta}).

\vspace{3cm}

{\bf Acknowledgements}:

 I thank Massimo Bianchi and Yassen Stanev for numerous illuminating discussions.

I thank the High Energy Section of the Abdus Salam ICTP, Trieste, where part of this work was done.

I thank Jurgen Fuchs, Christoph Schweigert, Johannes Walcher, and
Ilka Brunner for useful e-mail correspondence, clarifying their
works.

This work was supported in part by INFN, by the MIUR-COFIN
contract 2003-023852, by the EU contracts MRTN-CT-2004-503369 and
MRTN-CT-2004-512194, by the INTAS contract 03-516346 and by the
NATO grant PST.CLG.978785.

\newpage

\appendix

\section{Theta functions identities}
We start by reviewing some useful identities satisfied by Theta
functions \cite{DM}.

\bea\label{mumidd3}&& \theta\left[\matrix{\frac{a}{n_1}\cr
0\cr}\right](x_1,n_1\tau)\theta\left[\matrix{\frac{b}{n_2}\cr
0\cr}\right](x_2,n_2\tau) =\\ \nonumber
&&\sum_{\mu=0}^{n_1+n_2-1}\theta\left[\matrix{\frac{n_1\mu+a+b}{n_1+n_2}\cr
0\cr}\right](x_1+x_2,(n_1+n_2)\tau)\cdot\\ \nonumber
&&\theta\left[\matrix{\frac{n_1n_2\mu+n_2a-n_1b}{n_1n_2(n_1+n_2)}\cr
0\cr}\right](n_2x_1-n_1x_2,n_1n_2(n_1+n_2)\tau) \eea where
\be\label{thetab} \theta\left[\matrix{a\cr
b\cr}\right](x,\tau)=\sum_{n\in
Z}\exp(i\pi(n+a)^2\tau+2i\pi(n+a)(x+b)) \ee

Using the identity

\be\label{thetsum}
\sum_{\mu=0}^{n-1}\theta\left[\matrix{\frac{\mu+a}{n}\cr
0\cr}\right](nx,n^2\tau)=\theta\left[\matrix{a\cr
0\cr}\right](x,\tau) \ee

we can exploit (\ref{mumidd3}) for the case relevant to our analysis i.e.
$n_1=r_1n$ and $n_2=r_2n$ \bea\label{mumidd2}&&
\theta\left[\matrix{\frac{a}{r_1n}\cr
0\cr}\right](x_1,r_1n\tau)\theta\left[\matrix{\frac{b}{r_2n}\cr
0\cr}\right](x_2,r_2n\tau) =\\ \nonumber
&&\sum_{\mu=0}^{r_1+r_2-1}\theta\left[\matrix{\frac{r_1\mu}{r_1+r_2}+\frac{a+b}{(r_1+r_2)n}\cr
0\cr}\right](x_1+x_2,(r_1+r_2)n\tau)\cdot\\ \nonumber
&&\theta\left[\matrix{\frac{\mu}{r_1+r_2}+\frac{r_2a-r_1b}{r_1r_2(r_1+r_2)n}\cr
0\cr}\right](r_2x_1-r_1x_2,r_1r_2(r_1+r_2)n\tau) \eea

Let us explicitly write this formula for the most relevant for us
case: $n_1=n_2=n$, $r_1=r_2=1$

\bea\label{mumid} &&\theta\left[\matrix{\frac{a}{n}\cr
0\cr}\right](x_1,n\tau)\theta\left[\matrix{\frac{b}{n}\cr
0\cr}\right](x_2,n\tau)
=\\ \nonumber
&&\sum_{\mu=0}^1\theta\left[\matrix{\frac{\mu}{2}+\frac{a+b}{2n}\cr
0\cr}\right](x_1+x_2,2n\tau)
\theta\left[\matrix{\frac{\mu}{2}+\frac{a-b}{2n}\cr
0\cr}\right](x_1-x_2,2n\tau) \eea

\section{Other relevant identities}

Recall the identities: \bea\label{double}
&&\theta_3^2(\tau)-\theta_4^2(\tau)=2\theta_2^2(2\tau)\\ \nonumber
&&\theta_3^2(\tau)+\theta_4^2(\tau)=2\theta_3^2(2\tau)\\ \nonumber
&&\theta_3(\tau)\theta_4(\tau)=\theta_4^2(2\tau)\\ \nonumber
&&\theta_2^2(\tau)=2\theta_2(2\tau)\theta_3(2\tau) \eea From
(\ref{double}) we can derive another couple of useful identities:
\bea\label{23chan}
&&\theta_3(2\tau)\theta_2^2(\tau)=\theta_2(2\tau)(\theta_3^2(\tau)+\theta_4^2(\tau))\\
\nonumber
&&\theta_2(2\tau)\theta_2^2(\tau)=\theta_3(2\tau)(\theta_3^2(\tau)-\theta_4^2(\tau))\eea

 \bea\label{thetd} &&\Theta_{0,1}(z,\tau)=\theta_3(z,2\tau)\\
\nonumber &&\Theta_{1,1}(z,\tau)=\theta_2(z,2\tau) \eea

Let us also mention the following formulae.

\be \theta_1\left(\frac{1}{2},\tau\right)=\theta_2(0,\tau) \ee \be
\theta_2\left(\frac{1}{2},\tau\right)=0 \ee \be
\theta_3\left(\frac{1}{2},\tau\right)=\theta_4(0,\tau) \ee \be
\theta_4\left(\frac{1}{2},\tau\right)=\theta_3(0,\tau) \ee

\be
\theta_1\left(\frac{1}{4},\tau\right)=\theta_2\left(\frac{1}{4},\tau\right)=\theta_1\left(\frac{3}{4},\tau\right)=-\theta_2\left(\frac{3}{4},\tau\right)
\ee

\be
\theta_3\left(\frac{1}{4},\tau\right)=\theta_4\left(\frac{1}{4},\tau\right)=\theta_3\left(\frac{3}{4},\tau\right)=\theta_4\left(\frac{3}{4},\tau\right)
\ee

\be
\frac{\theta^2_3(\frac{1}{4},\tau)}{\theta^2_1(\frac{1}{4},\tau)}=\frac{\theta_3(0,2\tau)}{\theta_2(0,2\tau)}=
\frac{\theta_2^2(0,\tau)}{\theta_3^2(0,\tau)-\theta_4^2(0,\tau)} \ee

\section{Partition function of the $T^4/Z_4$ orbifold}
\be\label{otpf} Z=\frac{1}{4}Z_{\rm
lattice}\left|\frac{{\cal J}}{\eta^{12}}\right|^2+\sum_{r,s}' n_{r,s}|Z_{r,s}|^2 \ee where \be
Z_{\rm lattice}=(|\chi^{\rm SU(2)}_{1}|^2+|\chi^{\rm
SU(2)}_2|^2)^4=\frac{1}{4}\left(|\theta_3(0,\tau)|^4+|\theta_4(0,\tau)|^4+|\theta_2(0,\tau)|^4\right)^2
\ee and \be Z_{r,s}=\sum_{\alpha,\beta}c_{\alpha,\beta}
\frac{\theta^2\left[\matrix{\alpha\cr
\beta\cr}\right](0,\tau)}{\eta^6}\frac{\theta\left[\matrix{\alpha+\frac{r}{4}\cr
\beta+\frac{s}{4}\cr}\right](0,\tau)\theta\left[\matrix{\alpha-\frac{r}{4}\cr
\beta-\frac{s}{4}\cr}\right](0,\tau)}{\theta\left[\matrix{\frac{1}{2}+\frac{r}{4}\cr
\frac{1}{2}+\frac{s}{4}\cr}\right](0,\tau)\theta\left[\matrix{\frac{1}{2}-\frac{r}{4}\cr
\frac{1}{2}-\frac{s}{4}\cr}\right](0,\tau)} \ee Consider the Ramond
part. \be\label{zrs}
Z^{R}_{r,s}=\frac{\theta^2_2(0,\tau)}{\eta^6}\frac{\theta\left[\matrix{\frac{1}{2}+\frac{r}{4}\cr
0+\frac{s}{4}\cr}\right](0,\tau)\theta\left[\matrix{\frac{1}{2}-\frac{r}{4}\cr
0-\frac{s}{4}\cr}\right](0,\tau)}{\theta\left[\matrix{\frac{1}{2}+\frac{r}{4}\cr
\frac{1}{2}+\frac{s}{4}\cr}\right](0,\tau)\theta\left[\matrix{\frac{1}{2}-\frac{r}{4}\cr
\frac{1}{2}-\frac{s}{4}\cr}\right](0,\tau)} \ee \be\label{z01}
Z^{R}_{0,1}=-\theta_2^4(0,\tau)\frac{\theta^2_3(0,\tau)\theta_4^2(0,\tau)}{4\eta^{12}}
\ee \be\label{z02}
Z^{R}_{0,2}=0 \ee
\be\label{z03}
Z^{R}_{0,3}=-\theta_2^4(0,\tau)\frac{\theta^2_3(0,\tau)\theta_4^2(0,\tau)}{4\eta^{12}}
\ee \be\label{z20}
Z^{R}_{2,0}=\theta_2^4(0,\tau)\frac{\theta^4_3(0,\tau)}{4\eta^{12}}
\ee \be\label{z21}
Z^{R}_{2,1}=\theta_2^4(0,\tau)\frac{\theta^2_3(0,\tau)\theta_4^2(0,\tau)}{4\eta^{12}}
\ee \be\label{z22}
Z^{R}_{2,2}=\theta_2^4(0,\tau)\frac{\theta^4_4(0,\tau)}{4\eta^{12}}
\ee \be\label{z23}
Z^{R}_{2,3}=\theta_2^4(0,\tau)\frac{\theta^2_3(0,\tau)\theta_4^2(0,\tau)}{4\eta^{12}}
\ee \be\label{z10}
Z^{R}_{1,0}=\theta_2^4(0,\tau)\frac{\theta_3^4(0,\tau)+\theta^2_2(0,\tau)\theta_3^2(0,\tau)}{4\eta^{12}}
\ee \be\label{z12}
Z^{R}_{1,2}=-\theta_2^4(0,\tau)\frac{\theta_3^4(0,\tau)-\theta^2_2(0,\tau)\theta_3^2(0,\tau)}{4\eta^{12}}
\ee \be\label{z30}
Z^{R}_{3,0}=\theta_2^4(0,\tau)\frac{\theta_3^4(0,\tau)+\theta^2_2(0,\tau)\theta_3^2(0,\tau)}{4\eta^{12}}
\ee \be\label{z32}
Z^{R}_{3,2}=-\theta_2^4(0,\tau)\frac{\theta_3^4(0,\tau)-\theta^2_2(0,\tau)\theta_3^2(0,\tau)}{4\eta^{12}}
\ee \be\label{z11}
Z^{R}_{1,1}=\theta_2^4(0,\tau)\frac{\theta_4^4(0,\tau)+i\theta^2_2(0,\tau)\theta_4^2(0,\tau)}{4\eta^{12}}
\ee \be\label{z13}
Z^{R}_{1,3}=\theta_2^4(0,\tau)\frac{\theta_4^4(0,\tau)-i\theta^2_2(0,\tau)\theta_4^2(0,\tau)}{4\eta^{12}}
\ee \be\label{z31}
Z^{R}_{3,1}=\theta_2^4(0,\tau)\frac{\theta_4^4(0,\tau)-i\theta^2_2(0,\tau)\theta_4^2(0,\tau)}{4\eta^{12}}
\ee \be\label{z33}
Z^{R}_{3,3}=\theta_2^4(0,\tau)\frac{\theta_4^4(0,\tau)+i\theta^2_2(0,\tau)\theta_4^2(0,\tau)}{4\eta^{12}}
\ee
 The numbers $n_{r,s}$ are given by the following formulae:
$n_{0,s}=4\sin^4\frac{\pi s}{4}$, $n_{r,s}=n_{r,s+r}$,
$n_{r,s}=n_{s,4-r}$.

 Plugging all in (\ref{otpf}) we get (\ref{pfge}).

\section{Annulus partition functions for simple current extensions}

In this section we mainly follow \cite{Fuchs:1996dd}.

Let us start with the partition function:
 \be
Z=\sum_{\rm{orbits}\, Q(a)=0}|{\cal S}_a|\cdot |\sum_{J\in
G/{\cal S}_a}\chi_{Ja}|^2\ee
 where $G$ is a group of simple currents and
${\cal S}_a$ is the stabilizer of $a$. Denote by $G_a=G/{\cal S}_a$ the factor group
acting non-trivially on $a$. The order $|G_a|$ of $G_a$ is
$|G_a|=|G|/|{\cal S}_a|$.
 Let us write $|{\cal S}_a|$ as a sum of squares:
\be\label{msq} |{\cal S}_a|=\sum_i
(m_{a,i})^2
\ee
where $i$ labels the different primaries into
which $a$ gets resolved (usually
 $m_{a,i}$ has to be independent of $i$, but to keep track of
 the different sums we will keep the index $i$.)
 Corresponding to this definition we have
\be
 \tilde{\chi}_{a,i}=m_{a,i}\sum _{J\in G_a}\chi_{Ja}
\ee
 so that $\sum_i |\tilde{\chi}_{a,i}|^2=|{\cal S}_a|\cdot |\sum_{J\in
G/{\cal S}_a}\chi_{Ja}|^2$. The ansatz for the resolved modular
$S$ matrix suggested in \cite{Schellekens:1990xy} is
\be\label{schya}
\tilde{S}_{(a,i),(b,j)}=m_{a,i}m_{b,j}{|G_a||G_b|\over
|G|}S_{a,b}+\Gamma_{(a,i),(b,j)}\ee where $\Gamma_{(a,i),(b,j)}$
satisfies \be\label{nco} \sum_j\Gamma_{(a,i),(b,j)}m_{b,j}=0 . \ee
Now we derive the unitarity condition on
$\tilde{S}_{(a,i),(b,j)}$. Recall that $S_{a,b}$ satisfies
\be\label{scid} S_{Ja,b}=e^{2\pi i Q(b)}S_{a,b} \ee Computing
$\tilde{S}\tilde{S}^{\dagger}$ we produce four terms \be\label{1t}
P_{S,S}=\sum_{\rm{orbits}\,
Q(b)=0,j}m_{a,i}m_{b,j}^2m_{c,k}S_{a,b}S_{c,b}^*|G_a||G_b|^2|G_c|/|G|^2
\ee \be\label{2t} P_{S,\Gamma}= \sum_{\rm{orbits}\,
Q(b)=0,j}m_{a,i}m_{b,j}{|G_a||G_b|\over
|G|}S_{a,b}\Gamma_{(c,k),(b,j)}^* \ee \be\label{3t}
P_{\Gamma,S}=\sum_{\rm{orbits}\,
Q(b)=0,j}\Gamma_{(a,i),(b,j)}^{\star}m_{b,j}m_{c,k}{|G_b||G_c|\over
|G|}S_{c,b}^* \ee \be\label{4t} P_{\Gamma,
\Gamma}=\sum_{\rm{orbits}\,
Q(b)=0,j}\Gamma_{(a,i),(b,j)}\Gamma_{(c,k),(b,j)}^* . \ee We see
that, due to (\ref{nco}), (\ref{2t}) and (\ref{3t}) are $0$. Now
we evaluate (\ref{1t}). The sum over $j$ can be carried out using
(\ref{msq}): \be P_{S,S}=\sum_{\rm{orbits}\,
Q(b)=0}m_{a,i}m_{c,k}S_{a,b}S_{c,b}^*|G_a||G_b||G_c|/|G| . \ee The
sum over $b$ here runs over representatives of neutral orbits.
Using (\ref{scid}) we can extend it to sum over all values of $b$.
Using that $a$, $b$ and $c$ are neutral we get that $S_{a,b}$ and
$S_{c,b}$ are independent of the specific orbit representative
$S_{a,b}=S_{Ja,Kb}$ and $S_{c,b}=S_{c,Kb}$. Using this observation
we can write

\be P_{S,S}=\sum_{\rm{orbits}\, Q(b)=0}\sum_{J\in G_a,K\in
G_b}m_{a,i}m_{c,k}S_{Ja,Kb}S_{c,Kb}^*|G_c|/|G|
\ee
Again using
(\ref{scid}) we deduce that the sum over $a$ allows us to extend the sum
over $b$ from neutral orbits to all orbits. The sum over $a$
projects out the charged one. Now when we sum over all values of
$b$ we can use unitarity of $S$ and finally write \be
P_{S,S}={\delta_{ac}m_{a,i}m_{a,k}\over |{\cal S}_a|} .
\ee
We get that
unitarity imposes the following constraint on $\Gamma$
\be
\sum_{\rm{orbits}\,
Q(b)=0,j}\Gamma_{(a,i),(b,j)}\Gamma_{(c,k),(b,j)}^*=\delta_{ac}(\delta_{ik}-{m_{a,i}m_{a,k}\over
|{\cal S}_a|})
\ee
derived in \cite{Schellekens:1990xy}. Using the same tricks we
turn to the computation of the fusion coefficients.
\be\label{cofu} \tilde{{\cal
N}}_{(a,i),(c,k)}^{(d,e)}=\sum_{(b,j)}{\tilde{S}_{(a,i),(b,j)}\tilde{S}_{(c,k),(b,j)}\tilde{S}_{(b,j),(d,e)}^*\over
\tilde{S}_{(0),(b,j)}}
\ee
We know that the vacuum state has trivial
stabilizer. Therefore
\be\label{vac}\tilde{S}_{(0),(b,j)}=m_{b,j}|G_b|S_{0,b}
\ee
Inserting (\ref{schya}) and (\ref{vac}) in (\ref{cofu}) we obtain
the following eight terms:
\be\label{fc1}
P_{SSS}=\sum_{(\rm{orbits}\,
Q(b)=0,j)}{m_{a,i}m_{b,j}^2m_{c,k}m_{d,e}|G_a||G_b|^2|G_c||G_d|S_{a,b}S_{c,b}S_{b,d}^*\over
|G|^3S_{0,b}}\ee \be\label{fc2} P_{S\Gamma S}=\sum_{(\rm{orbits}\,
Q(b)=0,j)}{m_{a,i}m_{b,j}m_{d,e}|G_a||G_b||G_d|S_{a,b}\Gamma_{(c,k),(b,j)}S_{b,d}^*\over
|G|^2S_{0,b}}\ee \be\label{fc3} P_{\Gamma S
S}=\sum_{(\rm{orbits}\,
Q(b)=0,j)}{m_{c,k}m_{b,j}m_{d,e}|G_c||G_b||G_d|\Gamma_{(a,i),(b,j)}S_{c,b}S_{b,d}^*\over
|G|^2S_{0,b}}\ee\be\label{fc4} P_{\Gamma\Gamma
S}=\sum_{(\rm{orbits}\,
Q(b)=0,j)}{m_{d,e}|G_d|\Gamma_{(a,i),(b,j)}\Gamma_{(c,k),(b,j)}S_{b,d}^*\over
|G|S_{0,b}}\ee\be\label{fc5} P_{SS\Gamma}=\sum_{(\rm{orbits}\,
Q(b)=0,j)}{m_{a,i}m_{b,j}m_{c,k}|G_a||G_b||G_c|S_{a,b}S_{c,b}\Gamma_{(b,j),(d,e)}^*\over
|G|^2S_{0,b}}\ee \be\label{fc6}
P_{S\Gamma\Gamma}=\sum_{(\rm{orbits}\,
Q(b)=0,j)}{m_{a,i}|G_a|S_{a,b}\Gamma_{(c,k),(b,j)}\Gamma_{(b,j),(d,e)}^*\over
|G|S_{0,b}}\ee \be\label{fc7} P_{\Gamma
S\Gamma}=\sum_{(\rm{orbits}\,
Q(b)=0,j)}{m_{c,k}|G_c|\Gamma_{(a,i),(b,j)}S_{c,b}\Gamma_{(b,j),(d,e)}^*\over
|G|S_{0,b}}\ee\be\label{fc8}
P_{\Gamma\Gamma\Gamma}=\sum_{(\rm{orbits}\,
Q(b)=0,j)}{\Gamma_{(a,i),(b,j)}\Gamma_{(c,k),(b,j)}\Gamma_{(b,j),(d,e)}^*\over
m_{b,j}|G_b|S_{0,b}}\ee We see that thanks to (\ref{nco}) the
three terms containing two $S$ and one $\Gamma$, namely
$P_{S\Gamma S}$,$P_{\Gamma SS}$,$P_{SS\Gamma }$ given by
(\ref{fc2}), (\ref{fc3}) and (\ref{fc5}) correspondingly are zero.
Further simplifications occur when we consider the annulus amplitude:
\be\label{anamp} A_{(a,i),(d,e)}=\sum_{\rm{orbits}\,
Q(c)=0,k}\tilde{{\cal
N}}_{(a,i),(c,k)}^{(d,e)}\tilde{\chi}_{c,k}=\sum_{\rm{orbits}\,
Q(c)=0,k}\tilde{{\cal N}}_{(a,i),(c,k)}^{(d,e)}m_{c,k}\sum _{J\in
G_c}\chi_{Jc}\ee Now again due to (\ref{nco}) also (\ref{fc4}),
(\ref{fc6}) and (\ref{fc8}) have vanishing contribution. We see
that for the evaluation of the  annulus amplitude it is enough to
consider only the term (\ref{fc1}) and (\ref{fc7}). Using the
same tricks as in the check of unitarity we can easily compute
(\ref{fc1}). First using (\ref{msq}) we perform the sum over $j$
and obtain \be\label{fcc1} P_{SSS}=\sum_{\rm{orbits}\,
Q(b)=0}{m_{a,i}m_{c,k}m_{d,e}|G_a||G_b||G_c||G_d|S_{a,b}S_{c,b}S_{b,d}^*\over
|G|^2S_{0,b}}\ee
 Again
using neutrality of $a$, $c$ and $d$ we can extend the sum over
$b$ from the representatives of the neutral orbits to the whole
orbit, absorbing $|G_b|$, and afterwards using neutrality of $b$
absorbing $G_a$ in the sum of $a$ over orbit: \be\label{fcc22}
P_{SSS}=\sum_{\rm{orbits}\, Q(b)=0}\sum_{J\in G_a,K\in
G_b}{m_{a,i}m_{c,k}m_{d,e}|G_c||G_d|S_{Ja,Kb}S_{c,Kb}S_{Kb,d}^*\over
|G|^2S_{0,Kb}}\ee As before the sum over $a$ allows us to extend the
sum from the  neutral orbits to all orbits. So we get that the sum over
$b$ runs over all values. Using Verlinde formula for the
modular S-matrix we obtain: \be\label{fcc2} P_{SSS}=\sum_{J\in
G_a}{m_{a,i}m_{c,k}m_{d,e}|G_c||G_d|{\cal N}_{Ja,c}^d\over
|G|^2}\ee Inserting now (\ref{fc7}) and (\ref{fcc2}) in
(\ref{anamp}) and again using (\ref{msq}) finally we obtain:
\bea\label{anampp} A_{(a,i),(d,e)}&=&\sum_{\rm{orbits}\,
Q(c)=0}\sum_{J\in G_a}{m_{a,i}m_{d,e}|G_d|{\cal N}_{Ja,c}^d\over
|G|}\sum _{K\in G_c}\chi_{Kc}\nonumber\\ &+&\sum_{\rm{orbits}\,
Q(c)=0}\sum_{(\rm{orbits}\,
Q(b)=0,j)}{\Gamma_{(a,i),(b,j)}S_{c,b}\Gamma_{(b,j),(d,e)}^*\over
S_{0,b}}\sum _{K\in G_c}\chi_{Kc}\eea

It is easy to check that $\sum_{J\in G_a}{\cal N}_{Ja,c}^d$ is
independent on the orbit representative of $c$: \be \sum_{J\in
G_a}{\cal N}_{Ja,Kc}^d=\sum_{J\in G_a}{\cal N}_{Ja,c}^d\ee Also it
is easy to see that if $Q(a)=0$, and $Q(d)=0$, ${\cal
N}_{a,c}^d\neq 0$ only if also $Q(c)=0$. So without
changing the result we can sum over all $c$ in  (\ref{anamp}) and omit
the sum over $K$: \be \sum_{\rm{orbits}\, Q(c)=0}\sum_{J\in
G_a}{m_{a,i}m_{d,e}|G_d|{\cal N}_{Ja,c}^d\over |G|}\sum _{K\in
G_c}\chi_{Kc}=\sum_c\sum_{J\in G_a}{m_{a,i}m_{d,e}{\cal
N}_{Ja,c}^d\over |{\cal S}_d|}\chi_c\ee

Let us momentarily consider the case when we have no fixed points.
In this case the annulus amplitude is \be A_{a,d}
=\sum_c\sum_{J\in G}{\cal N}_{Ja,c}^d\chi_c\ee This result can be
easily interpreted. It means that in the simple current orbifold
theory the Cardy boundary states are given by \be |a\rangle_{\rm{orbifold}}={1\over
\sqrt{|G|}} \sum_{J\in G}|Ja\rangle\ee This is the expected
result. This was the starting point of the Recknagel-Schomerus
boundary states construction in \cite{Recknagel:1997sb}.

$\Gamma_{(a,i),(b,j)}$ also satisfies the condition
(\ref{scid}): \be\label{scidd}\Gamma_{J(a,i),(b,j)}=e^{2i\pi
Q(b)}\Gamma_{(a,i),(b,j)}\ee

 In
(\ref{anampp}) sums over $(b ,j)$ and $c$ run over neutral
representatives, but to evaluate this sum in practice, we can
use (\ref{scidd}) and the same tricks as before to extend this
sum over all values of $b$ and $c$. So for the second part
in (\ref{anampp}) we have: \bea\label{aaa} &&\sum_{\rm{orbits}\,
Q(c)=0}\sum_{(\rm{orbits}\,
Q(b)=0,j)}{\Gamma_{(a,i),(b,j)}S_{c,b}\Gamma_{(b,j),(d,e)}^*\over
S_{0,b}}\sum _{K\in G_c}\chi_{Kc}\nonumber\\
&& =\sum_c \sum_{b,j}\sum_{J\in G_a}
{\Gamma_{J(a,i),(b,j)}S_{c,b}\Gamma_{(b,j),(d,e)}^*\over
S_{0,b}|G_b||G_a|}\chi_c\eea

Finally let us take into account that in the case of Gepner
models, all the stabilizer equal $Z_2$ and the unitarity condition can be satisfied taking  $\Gamma_{(a,\psi),(b,\psi')}$ in the form \be
\Gamma_{(a,\psi),(b,\psi')}={|G_a||G_b|\over
|G|}\hat{S}_{ab}\psi\psi'\delta_{af}\delta_{bf}\ee where the resolving index $\psi$ takes two values $\pm$ and
$\hat{S}_{ab}$ is some unitary matrix. Putting this in (\ref{aaa})
for this part of the annulus amplitude we get: \be{1\over
|{\cal S}_d|}\psi\psi''\sum_c \sum_b\sum_{J\in G_a}
{\hat{S}_{Ja,b}S_{c,b}\hat{S}_{b,d}^*\over
S_{0,b}}\chi_c\delta_{af}\delta_{bf}\delta_{df}\ee This is the
correction required when fixed points are present.  Putting
all pieces together we have:
\bea\label{bsch}
&&A_{(a,i),(d,e)}=\sum_c\sum_{J\in
G}{m_{a,i}m_{d,e}{\cal N}_{Ja,c}^d\over |{\cal S}_a||{\cal S}_d|}\chi_c+ \\ \nonumber
&&{1\over
|{\cal S}_a||{\cal S}_d|}\psi\psi''\sum_c \sum_b\sum_{J\in G}
{\hat{S}_{Ja,b}S_{c,b}\hat{S}_{b,d}^*\over
S_{0,b}}\chi_c\delta_{af}\delta_{bf}\delta_{df}
\eea

 This formula
was found by Brunner and Schomerus in \cite{Brunner:2000nk}.

\end{document}